# Cold accretion flows and chemical bimodality of the Milky Way galaxy

**Masafumi Noguchi[1]**

[1]Astronomical Institute, Tohoku University, Aoba-ku, Sendai 980-8578, Japan

**Abundance of chemical elements in the stars provides important clues regarding galaxy formation. The most powerful diagnostics is the relative abundance of α-elements (O, Mg, Si, S, Ca, and Ti) with respect to iron (Fe), [α/Fe], each of which is produced by different kinds of supernovae. The existence of two distinct groups of stars in the solar neighbourhood, one with high [α/Fe] and another with low [α/Fe], suggests that the stars in the solar vicinity have two different origins.[1,2,3] However, the specific mechanism of the realization of this bimodality is unknown. Here, we show that the cold flow hypothesis recently proposed for the accretion process of primordial gas onto forming galaxies[4] predicts two episodes of star formation separated by a hiatus 6-7 Gyr ago and naturally explains the observed chemical bimodality. We found that the first phase of star formation that forms high [α/Fe] stars is caused by the 'genuine' cold flow, in which unheated primordial gas accretes to the galactic disk in a freefall fashion. The second episode of star formation that forms low [α/Fe] stars is sustained by much slower gas accretion as the once-heated gas gradually cools by radiation. The cold flow hypothesis can also explain the large-scale variation in the abundance pattern observed in the Milky Way galaxy[2] in terms of the spatial variation of gas accretion history.**

The chemical compositions of stars are a rich source of information regarding galaxy formation processes because they encode the chemical properties of the interstellar medium (ISM) from which stars form. Particularly widely used is the [α/Fe] vs. [Fe/H] diagram because α-elements and iron are produced and returned to the ISM with different timescales by two kinds of supernovae (SNe)[1,2,3]. The main provider of α-elements is Type II SNe that inject synthesized elements into ISM, about $10^7$ year after the stellar birth. Much debate exists regarding the progenitor of another Type Ia SNe.[5] Nevertheless, the broad consensus is that the ejecta from Type Ia SNe is rich in iron and mixed into ISM with a much longer timescale, of the order of 1 Gyr after stellar birth. Therefore, generally, ISM has a high [α/Fe] just after star formation starts because only Type II SNe contribute in the early phase of chemical enrichment. However, this ratio gradually decreases as iron begins to be supplied by retarded Type Ia SNe.

The Milky Way galaxy (MW) provides a unique testbed for the chemical evolution model of galaxies because its individual stars can be resolved and examined in detail, which is impossible for most external galaxies. One remarkable and enigmatic feature of the stars in the solar neighbourhood is that there are two well-separated groups on the [α/Fe] vs. [Fe/H] diagram[1,2,3]. One group is characterized by a high abundance ratio [α/Fe]. The other has relatively low [α/Fe]. One hypothesis proposed to explain this bimodality is the radial migration of stars in the MW disk.[6] If the stars born in an inner region with more enhanced α-elements moved outward to the solar position, e.g., by the action of bar gravitational force, we will observe two populations with different chemical compositions. However, the efficiency of this mechanism is unclear.[7] Another possibility is that the star formation in the solar vicinity did not continuously proceed but halted for a certain period by some reason, leading to two separated episodes of star formation.[8]

How the star formation proceeds is largely governed by how the primordial gas accretes onto the forming galaxy. The longstanding paradigm for accretion is that the gas that entered the dark matter halo is heated by a shock wave to a high temperature that provides sufficient pressure for the gas to maintain equilibrium within the halo gravitational field. Then, the gas cools by emitting radiation and gradually accretes to the forming galaxy at the center of the dark matter halo (shock-heating theory and hot-mode accretion).[9]

Recently, a new hypothesis has been put forward that a significant part of the primordial gas remains cold when it enters the halo, and reaches the galaxy in narrow streams almost in a freefall fashion (cold flow theory).[4,10] This cold-mode accretion is prevalent in low-mass galaxies but appears even in massive galaxies in their early evolution phase. This hypothesis is invoked to explain the red color of massive elliptical galaxies, in combination with the action of energy feedback from active galactic nuclei.[11] It also gives one possible interpretation for the high incidence of strongly star-forming galaxies at high redshifts.[12]

To examine the differences these two theories bring about in the growing process of MW-like galaxies, we constructed a simple model of disk galaxy evolution in which a growing dark matter halo gathers the surrounding primordial gas, which subsequently accretes to the disk component of the galaxy and forms stars.[13] ISM is treated as a cold gas component contained in the disk. The shock-heating model is constructed by assuming that the gas is heated to the virial temperature when entering the host halo and accretes to the disk plane with the timescale of radiative cooling (or with the freefall time if the cooling time is shorter). The cold-flow model is based on the Dekel-Birnboim model.[4] It provides information on when a stable shock wave appears in the growing halo and the gas accretion changes from cold- to hot-mode. The gas is made to accrete with the freefall time in the initial cold-flow phase but with the shorter of the cooling time and the freefall time after the shock develops. The present virial mass of the galaxy is assumed to be $1.2 \times 10^{12} \ M_\odot$.[14] For this mass, the cold-flow model produces a stellar disk of $6 \times 10^{10} \ M_\odot$ with a stellar scale length of 3.0 kpc in the present epoch, in broad agreement with the values estimated for the MW.[15,16]

Fig. 1 depicts the time evolution of the two models and the resulting [α/Fe] vs. [Fe/H] diagram for the solar neighbourhood defined as an annulus with 7 kpc $< R <$ 9 kpc, where $R$ is the galactocentric radius. The bottom panels show that the cold-flow theory predicts a clear chemical bimodality, whereas the shock-heating model produces a single sequence on the [α/Fe] vs. [Fe/H] diagram. The locations of the two stellar groups in the cold-flow model agree with the observed ones.[2]

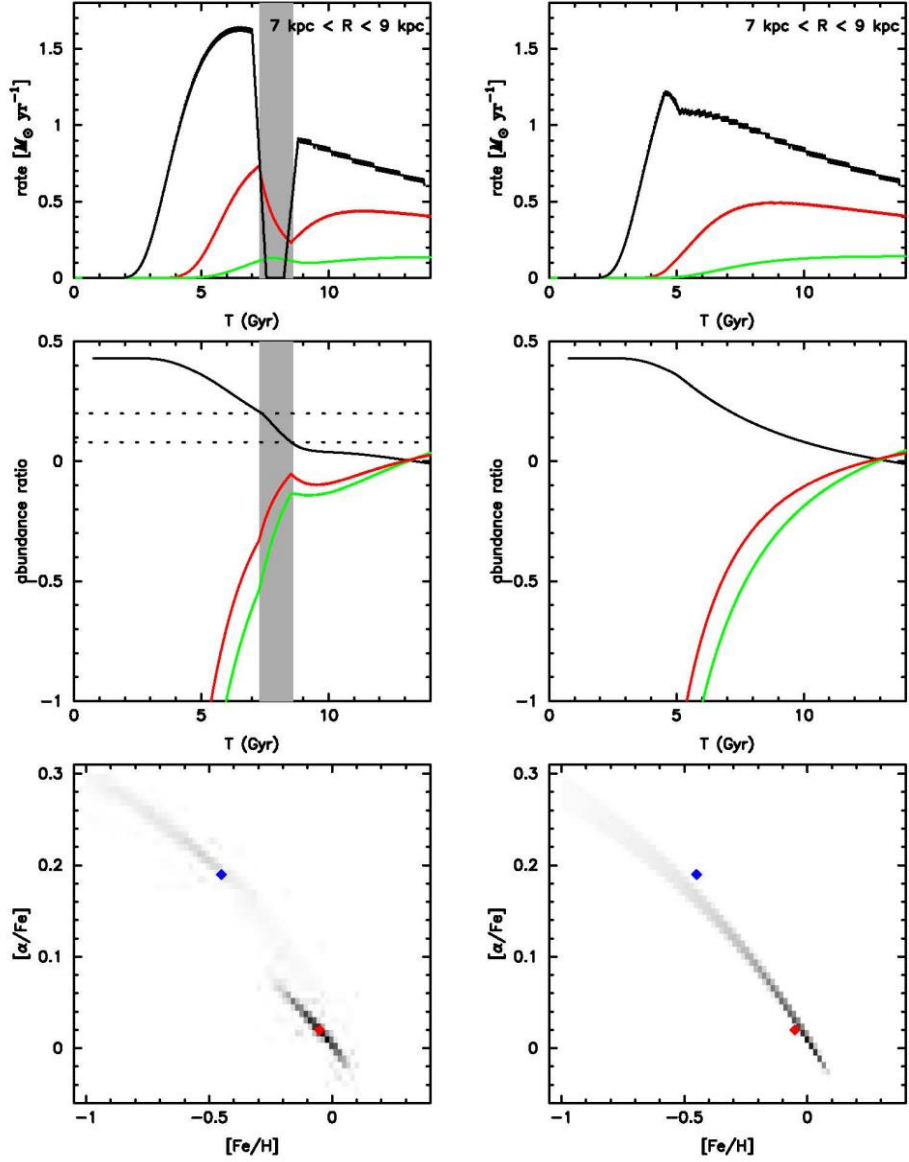

**Figure 1. Time evolution and the [α/Fe] vs. [Fe/H] diagram of the models.**
Left and right columns refer to the cold-flow and shock-heating models, respectively. Upper panels: the rate of gas accretion (black), star formation (red), and Type Ia SNe (green; number per century) are plotted against the cosmological time. Middle panels: Evolution of abundance ratios, [α/H] (red), [Fe/H] (green), and [α/Fe] (black). There is no accretion in the shaded period. Abundance ratio is the value relative to the solar value on the logarithmic scale. Solar abundances of $\left(\frac{\alpha}{H}\right) = 2.22 \times 10^{-2}$ and $\left(\frac{Fe}{H}\right) = 1.77 \times 10^{-3}$ are assumed.[25]

Bottom panels: Distribution of stars in the [α/Fe] vs. [Fe/H] diagram. The blue and red diamonds respectively indicate the approximate location of the peak of the high and low [α/Fe] sequences observed by APOGEE.[2]

The existence of a quiescent period about 6 Gyr ago is an essential event that contributes to creating the well-defined bimodality in the cold-flow model. In this period, star formation weakens, but Type Ia SNe explode frequently (upper panel, shaded region). During this period, [α/Fe] continues to decrease (middle panel), but only few stars are formed. The more fundamental cause is that gas accretion stops for a certain period during evolution.

How this dormant period of accretion arises in the cold-flow model is clarified in Fig.2. Here, $t_{ent}$ denotes the time at which the gas enters the dark matter halo. Therefore, Fig.2 shows the time at which the gas that accretes to the disk at the current time $T$ entered the halo. In the initial epoch ($t_{ent} < 4.3$ Gyr), the cooling time in the shock-heating model is shorter than the freefall time so that both models accrete gas in the same way. After $t_{ent} \sim 4.3$ Gyr, the gas continues to accrete with the freefall time in the cold-flow model, whereas gas accretion is delayed in the shock-heating model because the cooling time is longer than the freefall time. By $T \sim 7.3$ Gyr, the cold-flow model accretes the gas which accretes by $T \sim 8.6$ Gyr in the shock-heating model. At $t_{ent} \sim 6.3$ Gyr, the cold-mode ends, and the gas accretion is switched to the hot mode in the cold-flow model. Therefore, the gas that enters the halo after $t_{ent} \sim 6.3$ Gyr, should wait until $T \sim 8.6$ Gyr before it accretes. This change of the accretion mode gives rise to a period in which accretion is halted (shaded period in Fig.2). At $T \sim 8.6$ Gyr, accretion in the shock-heating model catches up with that in the cold-flow model, and the two models accrete the gas in the same way thereafter.

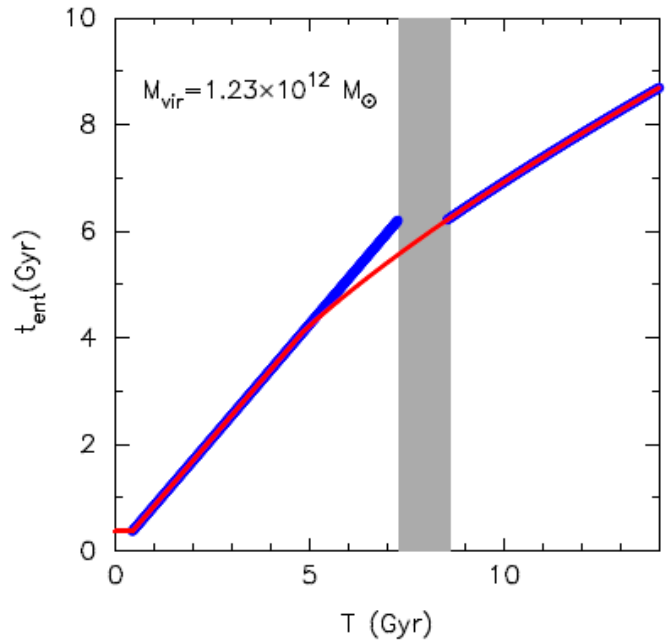

**Figure 2. Time of entering the dark matter halo as a function of arrival time to the disk.**
Blue and red lines indicate the cold-flow and shock-heating models, respectively. For example, at the present epoch ($T$=14 Gyr), the gas that entered the halo at $t_{ent} \sim 8.7$ Gyr is just accreting to the disk in both models. In the shaded period, no accretion occurs in the cold-flow model.

Fig. 1 details the evolution of the cold-flow model. The first gas accretion induces a brief period of active star formation, and α-elements start to be immediately released by Type II SNe. This star formation is prompt and begins to weaken before a significant amount of iron starts to be supplied by Type Ia SNe. Therefore, most stars formed during this epoch have high [α/Fe]. During the subsequent quiescent period in which star formation weakens, Type Ia SN explosions start to dominate. Since the enrichment of α-elements is overtaken by iron enrichment, [α/Fe] in ISM decreases further. When the second gas accretion starts ~2 Gyr later and the star formation is activated again, [α/Fe] stops changing rapidly and stays almost constant thereafter. The stars formed in this second phase have lower [α/Fe] ratios by ~ 0.2 dex than the first-generation stars. Paucity of stars with intermediate [α/Fe] values (between two dotted lines in the middle panel) creates a gap between the high and low [α/Fe] sequences in the [α/Fe] vs. [Fe/H] diagram.

The star formation history in the solar neighbourhood can be empirically deduced by matching the observed distribution on the [α/Fe] vs. [Fe/H] diagram with a prediction from an assumed star formation history using the stellar evolution library. Such an approach[8] reveals two phases of star formation separated by a sedation period about 7 Gyr ago in agreement with the prediction of the cold-flow model.

Recent large-scale surveys of individual stars such as GAIA-ESO[17], APOGEE[18], HARPS[19] extend the stellar abundance study to a sizable fraction of MW, reaching more than 10 kpc from the Sun. Interestingly, the prediction of the present cold-flow model can be compared with the observations for such a large-scale region of the galactic disk. Figure 3 shows the distribution of stars in the [α/Fe] vs. [Fe/H] diagram for three different regions: the inner disk (4 kpc < $R$ < 7 kpc), the solar neighborhood (7 kpc < $R$ < 9 kpc) and the outer disk (11 kpc < $R$ < 13 kpc).

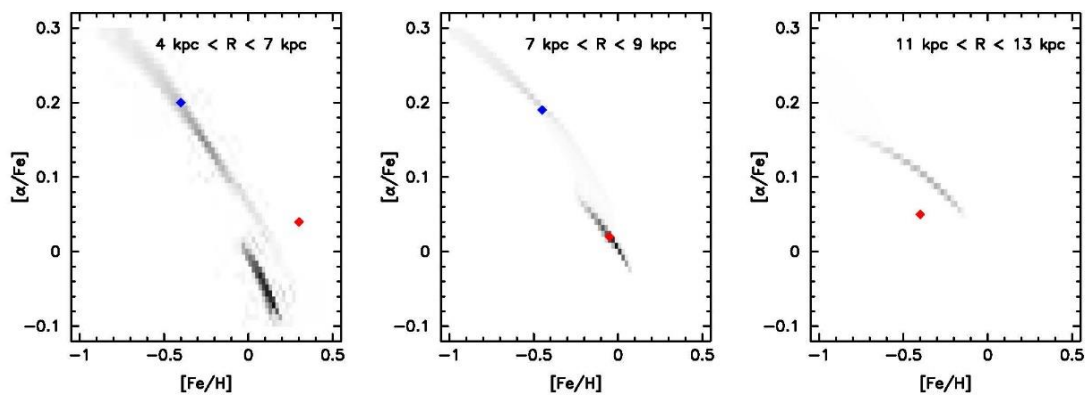

**Figure 3. Spatial variation in the [α/Fe] vs. [Fe/H] diagram.**
The distribution of stars for three different regions of the disk in the cold-flow model is shown. The number of stars decreases as we go outward. The blue and red diamonds in each panel respectively indicate the approximate location of the peak of the high and low [α/Fe] sequences observed by APOGEE for the corresponding region.[2]

The low [α/Fe] sequence increases its dominance over the high [α/Fe] one as we go outward because the second-phase accretion becomes more dominant compared with the first one (extended data, Fig.4). The high [α/Fe] sequence almost disappears in the outer disk. It is also noted that [α/Fe] of the low [α/Fe] sequence increases, whereas its [Fe/H] decreases as the galactocentric radius increases. This is because the outer disk is still in a younger phase of chemical evolution at present, making stars actively from a plentiful supply of ISM, so that the gas is not heavily polluted by iron yet. These trends are in broad agreement with the observation, though the observed [α/Fe] of the low [α/Fe] sequence shows no significant variation with radius.[2]

Thanks to recent advances in observational techniques, the galaxies in the Local Group and other nearby galaxies are becoming feasible targets for studying their star formation history by examining the color-magnitude diagram of their member stars. Because these nearby spiral galaxies span a wide range in mass and the cold-flow theory predicts different evolutions depending on the galaxy mass, such a nearfield "archaeology" provides powerful diagnostics for the theory.

M31, the largest galaxy in the Local Group, seems to have experienced two episodes of star formation separated by a break ~6 Gyr ago.[20] M33 and NGC300, which are several times less massive than MW, show a signature of inside-out formation and the star formation history seems monotonous at various places in each galaxy.[21,22] The cold-flow theory predicts that the quiescent period between the two star formation phases becomes longer for more massive galaxies, and an increasing number of stars are formed in the first phase. Interestingly, the break observed in M31 seems to be more prominent than the one inferred for MW. However, the relative dominance of the first- and second- generation stars appears to be independent on the radius, unlike in MW. According to the cold-flow model, MW has approximately the minimum mass for which a clear star formation gap is expected. Observations of M33 and NGC300[21,22] appear to support this prediction though the data still include a large uncertainty. It is hoped that future observations with higher sensitivity will clarify how star formation history and chemical evolution varies in galaxies with different masses.

## Methods

### The disk galaxy evolution model incorporating Type Ia supernovae

The MW model used in the present analysis was obtained by the code that divides a disk galaxy into a series of concentric annuli and calculates the time evolution of masses of gas and stars in each annulus under the specified gas accretion history and star formation law.[13] The original code treats the chemical enrichment of the interstellar gas attributed only to Type II SNe. For the present analysis, chemical enrichment due to Type Ia SNe is added. Specifically, α-elements of 3.177 $M_\odot$ and iron of 0.094 $M_\odot$ are added to the cold gas for one Type II SN. One Type Ia SN is assumed to return 1.38 $M_\odot$ of material to the cold gas, of which α-elements and iron occupy 0.438 $M_\odot$ and 0.74 $M_\odot$, respectively.[23] Type II SNe are assumed to explode instantaneously after the birth of progenitor stars. Type Ia SNe take a certain period after stellar birth before they explode. An observationally estimated delayed time distribution (DTD)[5] is introduced so that the rate for Type Ia SNe is given by

$$\text{DTD}(t) = 1 \times 10^{-3} \text{ SN Gyr}^{-1} \, M_{\odot}^{-1} \left(\frac{t}{\text{Gyr}}\right)^{-1.1}$$

, where $t$ denotes the time that elapsed since the progenitor formed. Considering the uncertainty in binary stellar evolution[24], we assume that DTD = 0 for $t < 0.3$ Gyr. The Type II SN feedback efficiency is taken to be $\varepsilon = 0.15$ instead of the original value 0.05 to match the model with MW, but this change does not influence the main conclusion of the present work. All other model parameters are the same as those adopted in Noguchi (2018).

**Data availability** The data that support the findings of this study are available from the corresponding author upon reasonable request.

---

# References

1. Adibekyan, V. Zh. et al. Chemical abundances of 1111 FGK stars from the HARPS GTO planet search program: Galactic stellar populations and planets. *Astron. Astrophys.* **545**, 32-46 (2012)

2. Anders, F. et al. Chemodynamics of the Milky Way I. The first year of APOGEE data. *Astron. Astrophys.* **564**, 115- (2014)

3. Hayden, M.R. et al. Chemical cartography with APOGEE: Metallicity distribution functions and the chemical structure of the Milky Way disk. *Astrophys. J.* **808**, 132-149 (2015)

4. Dekel, A. & Birnboim, Y. Galaxy bimodality due to cold flows and shock heating. *Mon. Not. R. Astron. Soc.* **368**, 2–20 (2006).

5. Maoz, D., Sharon, K., & Gal-Yam, A. The supernova delay time distribution in galaxy clusters and implications for type-Ia progenitors and metal enrichment. *Astrophys. J.* **722**, 1879-1894 (2010).

6. Schonrich, R. & Binney, J. Chemical evolution with radial mixing. *Mon. Not. R. Astron. Soc.* **396**, 203–222 (2009).

7. Minchev, I., Chiappini, C. & Martig, M. Chemodynamical evolution of the Milky Way disk I. The solar vicinity. *Astron. Astrophys.* **558**, 9-33 (2013)

8. Snaith, O. et al. Reconstructing the star formation history of the Milky Way disc(s) from chemical abundances. *Astron. Astrophys.* **578**, 87-115 (2015)

9. Rees, M.J. & Ostriker, J.P. Cooling, dynamics and fragmentation of massive gas clouds - Clues to the masses and radii of galaxies and clusters. *Mon. Not. R. Astron. Soc*. **179**, 541-559 (1977).

10. Fardal, M.A. et al. Cooling radiation and the Lyα luminosity of forming galaxies. *Astrophys. J.* **562**, 605-617(2001)

11. Cattaneo, A., Dekel, A., Devriendt, J., Guiderdoni, B. & Blaizot, J. Modelling the galaxy bimodality: Shutdown above a critical halo mass. *Mon. Not. R. Astron. Soc.* **370**, 1651–1665 (2006)

12. Dekel, A., et al. Cold streams in early massive hot haloes as the main mode of galaxy formation. *Nature*, **457,** 451-454 (2009)

13. Noguchi, M. Possible imprints of cold mode accretion on the present-day properties of disk galaxies. *Astrophys. J.*, **853**, 67-85 (2018)

14. Smith, M. C. et al.　The RAVE survey: constraining the local Galactic escape speed. *Mon. Not. R. Astron. Soc.* **379**, 755-772 (2007)

15. Bovy, J. & Rix, H.-W. A direct dynamical measurement of the Milky Way's disk surface density profile, disk scale length, and dark matter profile at 4 kpc $\lesssim$ r $\lesssim$ 9 kpc. *Astrophys. J.* **779**, 115-144 (2013)

16. Licquia, T.C. & Newman, J.A. Improved estimates of the Milky Way's stellar mass and star formation rate from hierarchical Bayesian meta-analysis. *Astrophys. J.* **806**, 96-115 (2015)

17. Gilmore, G. et al. The Gaia-ESO public spectroscopic survey. *Messenger*, **147**, 25-31 (2012)

18. Majewski, S.R. et al. The Apache Point Observatory Galactic Evolution Experiment (APOGEE). *Astron. J.* **154**, 94-139 (2017)

19. Mayor, M. et al. Setting new standards with HARPS. *Messenger*, **114**, 20-24 (2003)

20. Williams, B.F. et al. PHAT XIX. The ancient star formation history of the M31 disk. *Astrophys. J.* **846**, 145-180 (2017)

21. Williams, B.F., Dalcanton, J.J., Dolphin, A.E., Holtzman, J., & Sarajedini, A. The detection of inside-out disk growth in M33. *Astrophys. J.* **695**, L15-L19 (2009)

22. Gogarten, S.M. The Advanced Camera for Surveys Nearby Galaxy Survey Treasury. V. Radial star formation history of NGC 300. *Astrophys. J.* **712**, 858-874 (2010)

23. Ferreras, I. & Silk, J. Type Ia supernovae and the formation history of early-type galaxies. *Mon. Not. R. Astron. Soc.* **336**, 1181–1187 (2002)

24. Claeys, J.S.W. , Pols, O.R., Izzard, R.G., Vink, J. & Verbunt, F.W.M. Theoretical uncertainties of the Type Ia supernova rate. *Astron. Astrophys.* **563**, 83-106 (2014)

25. Asplund, M., Grevesse, N., Sauval, A. J., Scott, P. The chemical composition of the Sun. *Ann. Rev. Astron. Astrophys.* **47**, 481-522 (2009)

---

**Supplementary Information** is linked to the online version of the paper at www.nature.com/nature.

**Acknowledgments** We acknowledge I. Ferreras for helpful comments.

**EXTENDED DATA**

This is an extension of the Letter to Nature, aimed at providing further details, in support of the results reported in the main body of the Letter.

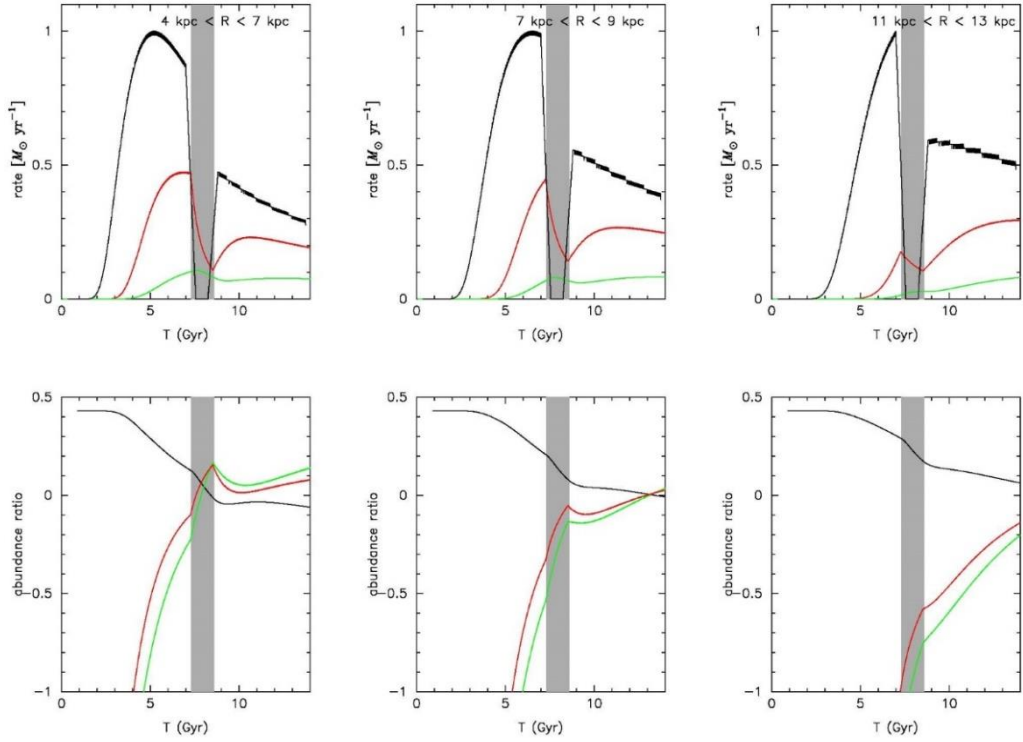

**Extended Data Figure 4. Time evolution of three different zones in the galactic disk of the cold-flow model.** These are the same diagrams as those in the upper and middle panels of Fig.1. In this figure, the maximum value of the accretion rate is normalized to unity in each panel for illustrative purpose. The star formation rate and Type Ia SN rate are scaled correspondingly. From left to right, time evolution in the inner disk, solar neighborhood, and outer disk is indicated. The dominance of the second accretion and associated star formation is noted to increase with the distance from the galactic center. On account of different timescales for enrichment, [α/Fe] decreases with time in all zones. Its value at the present epoch is higher at the outer radii as the star formation activity is in earlier phase owing to more efficient supply of cold gas, and the enrichment of α-elements is relatively more dominant than the retarded enrichment of iron by Type Ia SNe. [Fe/H] is also lower.